\documentclass[11pt]{article} 
 
\usepackage{bm,epsfig,amsmath} 
\usepackage{amssymb}
\usepackage{rotating}

\renewcommand\baselinestretch{1.12} 

\def\bp{{\mathbf p}} 
\def\br{{\mathbf r}} 
\def\bk{{\mathbf k}} 
\def\bu{{\mathbf u}} 
\def\bI{{\mathbf I}}

\def\rG{{\mathrm G}} 
\def\rM{{\mathrm M}} 
\def\rW{{\mathrm W}}

\def\rU{{\mathrm U}} 
\def\rY{{\mathrm Y}} 
 
\def\rg{{\mathrm g}}

\def\rtheta{{\mathrm\theta}}
\def\vphi{\varphi} 
 
\def\balpha{\mbox{$\bm{\alpha}$}}

\def\bnabla{\mbox{$\bm{\nabla}$}} 
 
\def\expp{{\mathrm e\,}} 
\def\prl{{\partial}}

\def\procref#1{  {\sf #1} }

\begin{document} 
 
\title{Relativistic wave and Green's functions for hydrogen--like ions} 
 
\author{Peter Koval\footnote{To whom correspondence should be 
addressed (kovalp@physik.uni-kassel.de)} \ and Stephan Fritzsche
\footnote{(s.fritzsche@physik.uni-kassel.de)} 
        \\ 
	\\ 
	\\ 
        Fachbereich Physik, Universit\"a{}t Kassel,  \\ 
        Heinrich--Plett--Str. 40, D--34132 Kassel, Germany 
	\\ 
	} 

\date{}

\maketitle

\thispagestyle{empty} 
\enlargethispage{0.0cm}

\begin{abstract} 
The \textsc{Greens} library is presented which provides a set of C++ procedures
for the computation of the (radial) Coulomb wave and Green's functions.
Both, the nonrelativistic as well as relativistic representations of these
functions are supported by the library. However, while the wave functions are 
implemented for all, the bound and free--electron states, the Green's functions
are provided only for bound--state energies $(E \,<\, 0$). Apart from the
Coulomb functions, moreover, the implementation of several special functions,
such as the Kummer and Whittaker functions of the first and second kind,
as well as a few utility procedures may help the user with the set--up and
evaluation of matrix elements.

\begin{center}
{\sf Koval~P and Fritzsche~S} 2003 {\it Comput.~Phys.~Commun.} {\bf 152} 191.
\end{center} 

\end{abstract} 

%

\newpage 
 
\textbf{\large PROGRAM SUMMARY} 
 
\bigskip

\textit{Title of program:} \textsc{Greens} 
 
\bigskip

\textit{Catalogue number:} To be assigned. 
 
\bigskip

\textit{Program obtainable from:} CPC Program Library, 
     Queen's University of Belfast, N.~Ireland.  
     Users may obtain the program also by down--loading a tar--file 
     from a home page at the University of Kassel\\    
     \texttt{http://www.physik.uni-kassel.de/$^\sim{}$kovalp/software/greens} 
 
\bigskip

\textit{Licensing provisions:} None. 
 
\bigskip

\textit{Computer for which the program is designed and has been tested:} 
     PC Pentium III, PC Athlon                                \newline 
     \textit{Installation:} University of Kassel (Germany).   \newline 
     \textit{Operating systems:} Linux 6.1+, SuSe Linux 7.3, SuSe Linux 8.0,
                                 Windows 98. 
      
\bigskip

\textit{Program language used:} C++. 
 
\bigskip

\textit{Memory required to execute with typical data:} 300 kB. 
 
\bigskip

\textit{No.\ of bits in a word:}  All real variables are of type
     \textbf{double} (i.e.\ 8 bytes long).
 
\bigskip

\textit{Distribution format:} Compressed tar file. 
 
\bigskip

\textit{CPC Program Library Subprograms required:} None.
   
\bigskip

 
\bigskip

\textit{Nature of the physical problem:}  \newline 
     In order to describe and understand the behaviour of hydrogen--like ions, 
     one often needs the Coulomb wave and Green's functions for the 
     evaluation of matrix elements. But although these functions have been
     known analytically for a long time and within different representations 
     [1,2], not so many implementations exist and allow for a simple access 
     to these functions. In practise, moreover, the application of the
     Coulomb functions is sometimes hampered due to \textit{numerical
     instabilities}.

\bigskip

\textit{Method of solution:}  \newline 
     The radial components of the Coulomb wave and Green's functions are
     implemented in position space, following the representation of Swainson 
     and Drake [2]. For the computation of these functions, however, use is 
     made of Kummer's functions of the first and second kind [3] which were
     implemented for a wide range of arguments. In addition, in order to 
     support the integration over the Coulomb functions, an adaptive 
     Gauss--Legendre quadrature has also been implemented
     within one and two dimensions.

\bigskip

\textit{Restrictions onto the complexity of the problem:}  \newline 
     As known for the \textit{hydrogen atom}, the Coulomb wave and Green's
     functions exhibit a rapid oscillation in their radial structure if either
     the principal quantum number or the (free--electron) energy increase.
     In the implementation of these wave functions, therefore, the 
     bound--state functions have been tested properly only up to the principal 
     quantum number $n \,\approx\, 20$, while the free--electron waves were 
     tested for the angular momentum quantum numbers $\kappa \,\le\,7$ and for
     all energies in the range $0 \, \ldots\, 10\,|E_{1s}| $.
     In the computation of the two--photon ionization cross sections $\sigma_2$,
     moreover, only the long--wavelength approximation 
     $(e^{\,i\mathbf{k\cdot r}} \,\approx\, 1)$ is considered both, within 
     the nonrelativistic and relativistic framework.

\bigskip

\textit{Unusual features of the program:} \newline 
     Access to the wave and Green's functions is given simply by means  of the
     \textsc{Greens} library which provides a set of C++ procedures. Apart from
     these Coulomb functions, however, \textsc{Greens} also supports the
     computation of several \textit{special functions} from mathematical 
     physics (see section 2.4) as well as of two--photon ionization cross 
     sections in long--wavelength approximation, i.e.\ for a very first
     application of the atomic Green's functions. Moreover, to facilitate 
     the integration over the radial functions, an adaptive Gauss--Legendre 
     quadrature has been also incorporated into the \textsc{Greens} library.

\bigskip

\textit{Typical running time:} 
     Time requirements critically depends on the quantum numbers and energies
     of the functions as well as on the requested accuracy in the case of a
     numerical integration. One value of the relativistic two--photon ionization
     cross section takes less or about one minute on a Pentium III 550 MHz 
     processor.

\bigskip 
 
{\it References:}   \newline 
[1] H.~A.~Bethe and E.~E.~Salpeter, \textit{Quantum Mechanics of One--and 
    Two--Electron Atoms},  (Kluwer Academic Publishers, 1977).
    \newline 
[2] R.~A.~Swainson and G.~W.~F.~Drake, J.~Phys.~A \textbf{24} 
    (1991) 95. 
    \newline  
[3] M.\ Abramowitz and I.~A.~Stegun, Eds., \textit{Handbook of Mathematical 
    Functions} (Dover, New York 1965).

\newpage 
{\large\bf LONG WRITE--UP} 
 
\bigskip 

\section{Introduction} 

From the early days of quantum mechanics on, the 'hydrogen atom' has served not
only as a well--known textbook problem but also as one of the fundamental 
models in the physics of atoms, molecules, and nuclei. When combined with the
(atomic) shell model, namely, the ---analytic--- solutions for the
hydrogen--like ions help understand most atomic processes in Nature, at least
qualitatively. For this reason also, the  'hydrogen atom' has found its way 
into quite different fields of physics including, for example, astro-- and 
plasma physics, quantum optics or even the search for more efficient x--ray 
lasers schemes. 

\medskip

Despite of the success of the \textit{hydrogen model}, however, the 
Coulomb problem is not always that simple to deal with, in particular, 
if a relativistic treatment is required. Therefore, various program tools 
have been developed over the years to help with either the analytic 
or numerical manipulation of the
Coulomb functions and their matrix elements. For the nonrelativistic
Coulomb problem, for example, the bound--electron states can be
obtained from the codes of Noble and Thompson \cite{Noble:84}, who applied
a continued fraction representation of the Whittaker 
functions, Bell and Scott \cite{Bell:80}, or simply by using the \textsc{Gnu}
Scientific Library \cite{GSL}. These functions are incorporated also into 
a recent library by Madsen and coworkers \cite{Madsen:99}, which has been 
designed to support the computation of the multipole matrix elements for 
circular and linear polarized light. --- Less attention, in contrast, 
has been paid to the \textit{relativistic} wave functions for which a 
\textsc{cpc} program is provided only by Salvat et al.\ \cite{Salvatetal:95}.
This program help integrate the radial equation for any spherical--symmetric
potential for both, the (one--particle) Schr\"o{}dinger and Dirac equations
and also provides separate procedures to compute the Coulomb wave functions.

\medskip

Apart from the bound and free--electron wave functions, however, the Coulomb
Green's functions play a similar important role, in particular, if the
interaction of atoms with external fields is to be studied. In second-- and
higher--order perturbation theory, for instance, these functions help to carry
out the summation over the complete spectrum in a rather efficient way. But
although different analytic representations are known for the Green's functions
\cite{Hostler:64a,Hostler:70,Mlodzki:84,Drake:91:2,Maquet:98}, until today,
there are almost no reliable codes freely available.

\medskip

Therefore, to facilitate the further application of the 'hydrogen atom' in 
different contexts, here we present the \textsc{Greens} library which provides
a set of C++ procedures for the computation of the Coulomb wave and Green's 
funtions. In \textsc{Greens}, these \textit{hydrogenic} functions are 
supported \textit{both}, within a nonrelativistic as well as 
relativistic framework. Beside of the various routines for the computation 
(of the radial parts) of these functions, however, we also supply the user
with a Gauss--Legendre quadrature and a set of special functions to simplify
the evaluation of matrix elements. --- But before we shall present details
about the organization of the \textsc{Greens} library, in the following
section, we first compile the basic formulas from the theory of
the 'hydrogen atom' with emphasize especially to those expressions, which
have been implemented explicitly. In section 3, later, the program structure 
will be discussed and how the library is to be distributed. This section 
also lists all user--relevant commands, although not much is said here about 
the underlying algorithms. In most cases, we followed the expressions 
from sections 2
but care has been taken in order to provide a \textit{reliable} code 
for a rather wide range of parameters which, sometimes, required quite
additional effort. In section 4, we explain how (easily) the hydrogenic 
wave and Green's functions can be accessed not only for a particular set 
of arguments but also for the computation of matrix elements. These examples 
may serve, therefore, also as a test bed for the installation of the code. 
Section 5, finally, gives a brief summary and an outlook into our future work.

\section{Theoretical Background} 
\label{theory}

Since the theory of the 'hydrogen atom' has been presented at quite many 
places before (see, for instance, the texts of Messiah \cite{Messiah:66}
and Drake \cite{DrakeHandBook}), we shall restrict ourselves to rather 
a short compilation of formulas, just enough in order to provide the basic 
notations and those expressions which are implemented in the code. 
In the next two subsections, therefore, we first recall the (analytic) form 
of the Coulomb wave and Green's functions while, in subsection 2.3, 
these functions are applied to calculate the two--photon ionizations 
cross sections for linear and circular polarized light. In all these 
subsections, the nonrelativistic and relativistic formulas are always presented
in turn of each other in order to display the similarities but also 
the differences in the numerical treatment of these functions. Subsection 2.4, 
moreover, provides reference to a few \textit{special functions} from 
mathematical physics, which frequently occur in the computations of the 
Coulomb wave or Green's functions and, hence, need to be part of the 
\textsc{Greens} library.

%
%
%
%
%
\subsection{Coulomb wave functions} 

\subsubsection{Nonrelativistic wave function} 

In a time--independent external field, the motion of a particle is described 
by the stationary Schr\"o{}dinger equation
\begin{eqnarray} 
\label{QME}    
   \left(\, \hat H\,(\br)-E \,\right) \,\psi\,(\br) & = & 0 
\end{eqnarray} 
which, obviously, is an eigenvalue equation for the total energy  $E$ of 
the particle. As known from the nonrelativistic Schr\"o{}dinger theory, 
the Hamiltonian $\hat H$ just includes the kinetic and potential energy 
of the particle and, thus, takes the form\,\footnote{Here and in the 
following, we use \textit{atomic units}
($m_e \,=\, \hbar \,=\, e^2/4\pi \epsilon_o \,=\, 1$) if not stated otherwise.}
\begin{eqnarray} 
\label{SH} 
   \hat H\,(\br) & = & -\frac{\nabla^{\,2}}{\,2} \:-\: \frac{Z}{r} \, 
\end{eqnarray} 
in the case of a (pure) Coulomb field of a nucleus with charge $Z$.
For such a spherical--symmetric potential, of course, Eq.\ (\ref{QME}) and 
the wave functions $\psi (\mathbf{r})$ can be separated
\begin{eqnarray}
\label{psinlm} 
   \psi_{nlm} (r,\theta,\vphi) & = & 
   \frac{P_{nl}\,(r)}{r} \: \rY_{lm}\,(\theta,\vphi) 
\end{eqnarray}
into a \textit{radial} and an \textit{angular} part where, in most practical
computations, the angular structure of the wave functions is often treated 
by means of the techniques from Racah's algebra \cite{Varshalovich:88}. 
In expression (\ref{psinlm}), $n$ and $l$ denote the principal
and orbital angular momentum quantum numbers, respectively, while $m$ 
describes the projection of the $z-$component of the orbital angular momentum
onto the quantization axis and is called the magnetic quantum number. 
The radial part of the wave function, $P_{nl}\,(r)/r$, is a solution of the 
\textit{radial} Schr\"o{}dinger equation
\begin{eqnarray} 
\label{nr-radial-equation}
   \left[\frac{1}{r^{\,2}} \,
         \frac{\prl}{\prl r}\left(r^2\frac{\prl }{\prl r} \right) 
         \: - \:
         \frac{l(l+1)}{r^{\,2}}\: + \:\frac{2Z}{r}\: + \:2E\right] \,
         \frac{P_{nl}(r)}{r}
   & = & 0 
\end{eqnarray}
which has (normalizable) \textit{physical solutions} for a discrete set of 
negative energies 
\begin{eqnarray}
\label{energy_non}
   E_n & = & -\frac{Z^2}{2\,n^2} \; < \; 0 , \qquad n \,=\, 1,\, 2,\, \ldots \, ,
\end{eqnarray}
the so--called \textit{bound} states, as well as for all positive energies 
$E\,>\,0$, i.e.\ the \textit{continuum} or \textit{free--electron} states. 
Both, the bound and continuum solutions of (\ref{nr-radial-equation}) can be 
represented in terms of a single Whittaker function of the first kind 
$M_{a,b}\,(z)$ 
\begin{eqnarray}
\label{radial-Pnl}
   P_{nl}\,(r) & = & C(n,l,Z)\: M_{n,\, l+\frac{1}{2}}\,(2Zr/n)  \\[0.2cm]
\label{radial-PEl}
   P_{El}\,(r) & = & C(E,l,Z)\: M_{i\sqrt{\frac{Z}{2E}},\, l+\frac{1}{2}}
                     (-2i\,\sqrt{2EZ}\, r)
\end{eqnarray} 
with real or complex arguments, and where $C(n,l,Z)$ and $C(E,l,Z)$,
respectively, denote the corresponding normalization factors. 
The Whittaker functions are closely related to the Kummer functions of first 
and second kind as we will discuss in subsection 2.2.1. In the standard theory,
moreover, the radial wave functions (\ref{radial-Pnl}) and (\ref{radial-PEl}) 
are often normalized due to 
\begin{eqnarray}
\label{nr-normalization}
   \int_0^{\infty} \, P_{nl}^2 \,(r) \, dr & = & 1,           \\[0.2cm]  
\label{nr-normalization-free}
   \int_0^{\infty} \, P_{El}^* \,(r) P_{E'l}\,(r) \, dr & = & \delta (E-E')\, ,  
\end{eqnarray}
in order to represent a single particle per bound state or 
\textit{per energy unit}, respectively, if particles in the continuum are
concerned.

\subsubsection{Relativistic wave functions} 

An eigenvalue equation analogue to (\ref{QME}) also applies,
if the motion of the particle is described within the relativistic theory.
For an electron with spin $s \,=\, 1/2$, however, then the Hamiltonian $\hat H$ 
needs to be replaced by the Dirac--Hamiltonian \cite{Messiah:66}
\begin{eqnarray} 
\label{DH} 
   \hat H_{\rm D}\,(\br) & = &  
   -i c \balpha \cdot \bnabla \,+\, \beta c^2 \,-\, \frac{Z}{r}  
\end{eqnarray}
which, apart from the kinetic and potential energy of the electron in the 
field of the nucleus, now also incorporates the rest energy of the electron as
well as energy contributions owing to its spin.
As in the nonrelativistic case, a separation of the wave function
\begin{eqnarray}
\label{greens_radial_spinor_n} 
   \psi_{n\kappa m}(\br) & = & 
   \frac{1}{r} \begin{pmatrix} 
                  P_{n\kappa}\,(r)\;\Omega_{\kappa m}\,(\theta,\vphi) \\[0.1cm] 
            i  \, Q_{n\kappa}\,(r)\;\Omega_{-\kappa m}\,(\theta,\vphi) 
               \end{pmatrix} 
\end{eqnarray} 
into a radial and angular part is possible for any spherical--symmetric 
potential, where the two radial functions $P_{n\kappa}\,(r)$ and 
$Q_{n\kappa}\,(r)$ are often called the \textit{large} and 
\textit{small} components. These two functions also form a radial spinor 
$\begin{pmatrix} P_{n\kappa}\,(r)\\[-0.1cm] Q_{n\kappa}\,(r)\end{pmatrix}$  
and have to be obtained as solutions of the first--order, coupled equations 
\cite{Drake:91:1}
\begin{eqnarray} 
\label{radial-Dirac-equation-P}
   \left[-\frac{Z}{r}-E \right]\frac{P_{n\kappa}\,(r)}{r} 
   \;+\; \left[\frac{\kappa}{\alpha r}
               \,-\, \frac{1}{\alpha r}\frac{\prl}{\prl r}r \right] \,
         \frac{Q_{n\kappa}\,(r)}{r} & = & 0 
         \\[0.2cm] 
\label{radial-Dirac-equation-Q}
   \left[\frac{1}{\alpha r}\frac{\prl}{\prl r} r 
   \,+\, \frac{\kappa}{\alpha r} \right] \,
   \frac{P_{n\kappa}\,(r)}{r} 
   \;-\; 
   \left[\frac{2}{\alpha^2}+\frac{Z}{r} \,+\, E \right]\,
         \frac{Q_{n\kappa}\,(r)}{r} & = & 0 \,  
\end{eqnarray} 
where, however, the (total) energy $E$ is taken here to represent the energy 
of the electron without its rest energy\footnote{In atomic units, the speed of
light $ c \,=\, 1/\alpha$ is the inverse of the fine--structure constant.} 
$c^{\,2}$, similar to 
Eq.\ (\ref{nr-radial-equation}) in the Schr\"o{}dinger theory. In Eqs.\ 
(\ref{greens_radial_spinor_n}--\ref{radial-Dirac-equation-Q}), moreover,
$\kappa \,=\, \pm\, (j+1/2)\,$ for $\,l \,=\, j \pm 1/2$ is called the 
\textit{relativistic} angular momentum quantum number and carries information
about both, the total angular momentum $j$ as well as the parity $(-1)^l$ of the
wave function. Again, (normalizable) \textit{physical solutions}
to the Dirac operator (\ref{DH}) can be found for a discrete set of negative 
energies 
\begin{eqnarray}
\label{energy_rel} 
   E_{n \kappa} & = & {\alpha^{-2} 
   \left[1+\left( \frac{\alpha Z}{n-\kappa+\sqrt{\kappa^2-\alpha^2 Z^2}}
           \right)^2 \right]^{-\frac{1}{2}}}
   \:-\: \alpha^{-2} \;<\; 0 \: , \nonumber  \\[0.2cm]
   &   & \hspace*{2.5cm}
   n \:=\: 1,\, 2,\, \ldots; \; 
   \kappa \:=\: -n,\, \ldots,\, n-1, \; \kappa \:\ne\: 0
\end{eqnarray}
and for all positive energies $E\,\ge\,0$ as well as for the (negative) 
energies $ E \,\le\, -2c^{\,2} $. The two latter ---contineous--- parts 
of the spectrum are also called the \textit{positive} and \textit{negative} 
continuum whereby the negative branch, in particular, requires some
re--intepretation of the theory (in terms of positron states, for example)
and often introduces additional complications in the treatment of 
many--electron systems. When compared with the nonrelativistic energies
(\ref{energy_non}), however, the degeneracy of the (relativistic) energies 
(\ref{energy_rel}) is partially resolved and now depends on both, the  
principal quantum number $n$ and the relativistic quantum number $\kappa$.

\medskip

Explicit representation of the bound and free--electron solutions of
Eqs.\ (\ref{radial-Dirac-equation-P}--\ref{radial-Dirac-equation-Q})
are known from the literature (cf.\ \cite{Drake:91:1,Eichler}) but 
typically result in rather lengthy expressions. For the bound states, 
for example, the two radial components are given by 
\begin{eqnarray}
\label{P-component} 
   P_{n\kappa}\,(r) & = & 
   C_P(n,\kappa,Z)\,\,r^s\expp^{-qr} \,
   \left[(-n+|\kappa|) \, \rM(-n+|\kappa|+1,\, 2s+1;\, 2qr) \right.  
   \nonumber \\  &  & \hspace*{3.9cm}
   - \, \left. \left(\kappa-Z\,q^{-1}\right) \,
   \rM(-n+|\kappa|,\, 2s+1;\, 2qr) \right]
   \\[0.2cm]
\label{Q-component} 
   Q_{n\kappa}\,(r) & = & 
   C_Q(n,\kappa,Z)\,\,r^s\expp^{-qr}\,
   \left[-(-n+|\kappa|) \, \rM(-n+|\kappa|+1,\, 2s+1;\, 2qr) \right.
   \nonumber \\  &  & \hspace*{3.9cm}
   -\left. \left(\kappa-Z\,q^{-1}\right) \,
   \rM(-n+|\kappa|,\, 2s+1;\, 2qr)\right]
\end{eqnarray}
where $\rM(a,b;z)$ is the Kummer function of the first kind,
$ s \,=\, \sqrt{\kappa^2-(\alpha Z)^2}$, and 
$ q \,=\, Z\,[(\alpha Z)^2 \,+\, (n-|\kappa|+s^2)]^{-\frac{1}{2}}$,
while even more elaborate expressions arise for the free--electron states
\cite{Eichler}. Similiar to (\ref{nr-normalization}) and 
(\ref{nr-normalization-free}),the bound and free--electron radial 
wave functions can be normalized also due to
\begin{eqnarray}
\label{r-normalization}
   \int_0^{\infty}\left( P_{n\kappa}^{\,2}\,(r) \,+\,  
                         Q_{n\kappa}^{\,2}\,(r) \right) \, dr 
   & = & 1,  \\[0.2cm]
   \int_0^{\infty}\left( P_{E\kappa}\,(r) P_{E'\kappa}\,(r) \,+\,   
                    Q_{E\kappa}\,(r) Q_{E'\kappa}\,(r) \right) \, dr 
   & = & \delta (E-E') 
\end{eqnarray}
to represent one electron per bound state or per energy unit, respectively.

\medskip

In the \textsc{Greens} library, the radial functions of the bound and
free--electron states can be accessed by means of the two library procedures
\procref{greens\_{}radial\_{}orbital()} and 
\procref{greens\_{}radial\_{}spinor()} in the nonrelativistic and relativistic
case, respectively; for further details, see section 3.

\subsection{Coulomb Green's functions} 

Apart from the wave functions, which describe the electron in particular 
quantum states, one often needs a summation over all (unoccupied) states, 
especially, if parts of the atomic interaction are treated as a perturbation. 
A full summation is required in second-- and higher--order perturbation theory,
for instance, if the behaviour of the atom is studied in a --- not too weak --- 
radiation field or in the presence of external electric or magnetic fields.
Although, in principle, it appears straightforward to carry out such a 
summation explicitly, the large number of terms and the need of 
\textit{free--free} matrix elements may hamper such an approach.
Instead, the use of Green's functions \cite{MorseFeshbach}
\begin{eqnarray}
\label{greens-expansion}
   \rG_E(\br,\br') & = &
   \sum\mkern-26mu\int_{\nu} \; 
   \frac{|\psi_{\nu}\,(\br)\rangle \, \langle \psi_{\nu}\,(\br') |}{E_{\nu}-E},
\end{eqnarray}
often provides a much simpler access to the \textit{spectrum} of the atom 
and, hence, to a perturbative treatment of atomic processes. In the following,
therefore, we first recall a representation of the radial Coulomb Green's 
functions as appropriate for numerical computations. The application of these 
functions in the computation of two--photon ionization cross sections 
$\sigma_2$ for hydrogen--like is discussed later in 
subsection~\ref{s:twophotonme}.

\subsubsection{Nonrelativistic Green's function} 

Analogue to the wave functions (\ref{psinlm}), the Coulomb Green's functions
$ \rG_E(\br,\br') $ are obtained as solutions of a linear equation
\begin{eqnarray}
\label{CoulombGreensEquation} 
   (\hat H(\br)-E)\, \rG_E(\br,\br') & = & \delta(\br-\br') \,  
\end{eqnarray} 
with the same Schr\"o{}dinger operator as in (\ref{QME}) but for an additional 
$\delta-$like inhomogenity on the right--hand side, which allows for solutions 
for any arbitrary $E$. For a spherical--symmetric potential, again, 
this equation can be separated into a radial and angular part by using 
the ansatz
\begin{eqnarray} 
\label{CoulombGreensNonSplit} 
   \rG_E(\br,\br') & = &  \sum_{lm}  \frac{\rg_{El}\,(r,r')}{rr'} \:
   \rY_{lm}(\theta,\vphi) \, \rY_{lm}^*(\theta',\vphi') 
\end{eqnarray}
for the Green's function in spherical coordinates. By substituting ansatz 
(\ref{CoulombGreensNonSplit}) into Eq.\ (\ref{CoulombGreensEquation}), 
one easily shows that the \textit{radial} Green's function $\rg_{El}\,(r,\,r')$,
which just depends on the energy $E$ and the orbital angular momentum $l$, 
must satisfy the equation
\begin{eqnarray*}
   \left[\frac{1}{r^{\,2}}\frac{\prl}{\prl r}\, 
   \left( r^2\frac{\prl}{\prl r} \right)
   \,-\, \frac{l(l+1)}{r^{\,2}} \,+\, \frac{2Z}{r} \,+\, 2E\right] 
   \frac{\rg_{El}\,(r,\,r')}{rr'}
   & = & -2\: \frac{\delta(r-r')}{rr'} \, . 
\end{eqnarray*} 
Solutions to this single equation can be determined by taking a proper 
superposition of the regular and irregular solutions (near the origin) of 
Schr\"o{}dinger's equation (\ref{nr-radial-equation}). An explicit 
representation for the radial Green's function reads as \cite{Drake:91:2}
\begin{eqnarray} 
\label{radial_Green_function_algebraic} 
   \rg_{El}\,(r,\,r') & = &
   \frac{\Gamma(l+1-\tau)}{x \, \Gamma(2l+2)} \:
   \rM_{\tau,\,l+\frac{1}{2}}\,(2x\, r_<) \:
   \rW_{\tau,\,l+\frac{1}{2}}\,(2x\, r_>) \, ,  
\end{eqnarray} 
where $x \,=\, (-2E)^{1/2},$ $\tau \,=\, \displaystyle\frac{Z}{x}$,
and where $r_{\,>}\,=\,\max(r,r')$ and $r_{\,<}\,=\,\min(r,r')$ refer to the 
\textit{larger} and \textit{smaller} value of the two radial coordinates, 
respectively. In this representation, moreover, $\rM_{\,a,b}\,(z)$ and 
$\rW_{\,a,b}\,(z)$ denote the two Whittaker functions of the first and second 
kind which can be expressed also in terms of the Kummer functions $\rM(a,b;z)$ 
and $\rU(a,b;z)$ of the corresponding kinds \cite{Abramowitz} 
\begin{eqnarray} 
\label{WhittakerM}
   \rM_{\,a,b}\,(z) & = & z^{\,b+\frac{1}{2}} \,
                          \expp^{-z/2}\:\rM(b-a+1/2, 2b+1; z) \, ,
   \\[0.1cm] 
\label{WhittakerW}
   \rW_{\,a,b}\,(z) & = & z^{\,b+\frac{1}{2}} \,
                          \expp^{-z/2}\:\rU(b-a+1/2, 2b+1; z)\, . 
\end{eqnarray}
In practise, the two \textit{Kummer functions} are used more frequently 
(than the Whittaker functions) in the mathematical literature and in various 
program libraries since $M(a,b;z)$ is closely related to the hypergeometric 
series and since the Kummer function $U(a,b;z)$ of the second kind can 
be expressed in terms of $M(a,b;z)$. In addition, several improved algorithms 
have been worked out recently in order to calculate the regular Kummer 
function $M(a,b;z)$ more efficiently, see section \ref{s:spec-funct} for 
further details.

\subsubsection{Relativistic Green's function} 

Of course, the relativistic Coulomb Green's function must refer to the Dirac 
Hamiltonian (\ref{DH}) and, hence, is given by a $4 \times 4$--matrix which 
satisfies the equation
\begin{eqnarray*} 
\label{CoulombDiracGreensEquation} 
   \left( \hat H_{\rm D} (\br)-E - c^{\,2}\, \right) \, \rG_E(\br,\br') 
   & = & \delta(\br-\br') \: \bI_{\,4} \, ,
\end{eqnarray*}
where $\bI_{\,4}$ denotes the $4\times 4$ unit--matrix and where, as for the 
wave functions from Eqs.\ 
(\ref{radial-Dirac-equation-P}--\ref{radial-Dirac-equation-Q}), the rest 
energy $c^{\,2}$ has not been incorporated into the (total) energy $E$.
Solutions to this equation are known again from the literature for a 
\textit{radial--angular} representation of the Coulomb Green's function 
\cite{Drake:91:2}
\begin{small}
\begin{eqnarray} 
\label{greens_radial_Green_matrix} 
   \rG_E(\br,\br') & = & \sum\limits_{\kappa m}\frac{1}{rr'} 
   \begin{pmatrix} 
      \rg_{E\kappa}^{\,LL}\,(r,\,r')\: 
      \Omega_{\kappa m}\,(\br)\, \Omega_{\kappa m}^{\dagger}(\br') & 
      -i\,\rg_{E\kappa}^{\,LS}\,(r,\,r')\: 
      \Omega_{\kappa m}\,(\br)\, \Omega_{-\kappa m}^{\dagger}(\br')\: 
      \\[0.25cm] 
      \,i\,\rg_{E\kappa}^{\,SL}\,(r,\,r')\: 
      \Omega_{-\kappa m}\,(\br)\,\Omega_{\kappa m}^{\dagger}(\br') & 
      \rg_{E\kappa}^{\,SS}\,(r,\,r')\: 
      \Omega_{-\kappa m}\,(\br)\,\Omega_{-\kappa m}^{\dagger}(\br') 
   \end{pmatrix} \, ,
\end{eqnarray} 
\end{small}

where the \textit{radial part} 
$\begin{pmatrix}
    \,\rg_{E\kappa}^{\,LL}\,(r, r') & \rg_{E\kappa}^{\,LS}\,(r, r') \\[0.2cm] 
    \,\rg_{E\kappa}^{\,SL}\,(r,r') & \rg_{E\kappa}^{\,SS}\,(r,r')\, 
 \end{pmatrix} /rr' $ of this function is now a $2\times 2$--matrix which must 
 satisfy the matrix equation  
\begin{eqnarray*} 
   \begin{pmatrix} 
   \left[\displaystyle - \frac{Z}{r} - E \right]                   &  
   \left[\displaystyle \frac{\kappa}{\alpha r} 
                  \,-\,\frac{1}{\alpha r}\frac{\prl}{\prl r}r \right]
   \\[0.55cm] 
   \left[\displaystyle \frac{1}{\alpha r}\frac{\prl}{\prl r} r 
         + \frac{\kappa }{\alpha r} \right]                     & 
   \left[\displaystyle -\frac{2}{\alpha^2}
         -\frac{Z}{r} - E \right] 
   \end{pmatrix} 
   \:\frac{1}{rr'}\: 
   \begin{pmatrix}\rg_{E\kappa}^{\,LL}\,(r, r') & 
                  \rg_{E\kappa}^{\,LS}\,(r, r') \\[0.35cm] 
                  \rg_{E\kappa}^{\,SL}\,(r, r') & 
                  \rg_{E\kappa}^{\,SS}\,(r, r') 
   \end{pmatrix} 
   & = & \frac{\delta(r-r')}{rr'}\: \bI_{\,2} \, .
\end{eqnarray*}
In this representation of the Green's function, we make use of the two 
superscripts
$T$ and $T'$ to denote the individual components in the $2\times 2$ radial Green's
matrix. They may take both the values $T \,=\, [L,S]$ to refer to either the
\textit{large} or \textit{small} components, when multiplied 
with a wave function spinor (\ref{greens_radial_spinor_n}). An explicit 
representation of the (four) components $\rg^{\,TT'}_{E\kappa}\,(r, \, r')$ of 
the radial Green's function is found by Swainson and  Drake \cite{Drake:91:2} 
\begin{small}
\begin{eqnarray}
\label{radial-green-matrix}
   &   & \hspace*{-1.5cm}
   \begin{pmatrix} \rg_{E\kappa}^{\,LL} & \rg_{E\kappa}^{\,LS} \\[0.2cm] 
                   \rg_{E\kappa}^{\,SL} & \rg_{E\kappa}^{\,SS} 
   \end{pmatrix} 
   \nonumber \\[0.3cm]
   & = & \frac{1}{(1-X^2)^2} \:
   \begin{pmatrix}
                   h^{11}\,-\,X(h^{12}+h^{21})\,+\,X^2 h^{22} & 
                   -X(h^{11}+h^{22})\,+\,h^{12}\,+\,X^2h^{21}  \\[0.2cm]
                   -X(h^{11}+h^{22})\,+\,X^2h^{12}\,+\,h^{21} & 
                   X^2\,h^{11}\,-\,X(h^{12}+h^{21})\,+\,h^{22}  
   \end{pmatrix},
\end{eqnarray}
with 
\begin{eqnarray}
\label{h11}
   h^{11}\,(r,\, r') & = & 
   \frac{(1-X^2)((E\alpha^2+1) \,\kappa\gamma^{-1}+1)}{2\omega} \,
   \frac{\Gamma(\gamma+1-\nu)}{\Gamma(2\gamma+2)}  \,
   \rM_{\nu,\,\gamma+\frac{1}{2}}(2\omega\, r_{<}) \,
   \rW_{\nu,\,\gamma+\frac{1}{2}}(2\omega\, r_{>}) \, 
   \\[0.2cm]
   h^{22}\,(r,\, r') & = &
   \frac{(1-X^2)((E\alpha^2+1) \, \kappa\gamma^{-1}-1)}{2\omega} \,
   \frac{\Gamma(\gamma-\nu)}{\Gamma(2\gamma)}      \,
   \rM_{\nu,\,\gamma-\frac{1}{2}}(2\omega\, r_{<}) \,
   \rW_{\nu,\,\gamma-\frac{1}{2}}(2\omega\, r_{>}) \, 
   \\[0.2cm]
\label{h21}
   h^{21}\,(r,\, r') & = & h^{12}\,(r',\,r) \nonumber \\[0.2cm]
   & = &
   \frac{(1-X^2)\,\Gamma(\gamma+1-\nu) \, \alpha \gamma^{-1} 
        }{2 \,\Gamma(2\gamma+2)}  \;
   \left[2\gamma(2\gamma+1)\,\rtheta(r'-r)\,
                  \rM_{\nu,\,\gamma-\frac{1}{2}}(2\omega r)\,
                  \rW_{\nu,\,\gamma+\frac{1}{2}}(2\omega r') \right.
   \nonumber  \\[0.2cm]
   &   & \hspace*{4.5cm}
   -\left. (\nu+\gamma)\,\rtheta(r-r')\,
            \rW_{\nu,\gamma-\frac{1}{2}}(2\omega r)\,
            \rM_{\nu,\gamma+\frac{1}{2}}(2\omega r') \right] \: 
\end{eqnarray}
\end{small}
and
\begin{eqnarray*}
   X & = & {(-\kappa+\gamma) (\alpha Z)^{-1}} \, ,      \quad \qquad \qquad
   \gamma \; = \; ({\kappa^2 - \alpha^2 Z^2})^{1/2} \, ,      \\[0.1cm]
   \omega & = & \alpha^{-1}(1-(E\alpha^{2}+1)^2)^{1/2} \, ,   \qquad
   \nu \; = \; { Z (E\alpha^2+1) \omega^{-1}} \, ,
\end{eqnarray*}
and where $\theta(x)$ denotes the Heaviside function. 

\medskip

In the \textsc{Greens} library, we provide the two procedures 
\procref{greens\_radial\_function()} and \newline
\procref{greens\_radial\_matrix()} which support the computation of the
radial functions (\ref{radial_Green_function_algebraic}) and 
(\ref{radial-green-matrix}) for any proper set of parameters.

\subsection{Two--photon transition amplitudes and ionization cross sections} 
\label{s:twophotonme} 
 
The Green's function (\ref{CoulombGreensNonSplit}) and 
(\ref{greens_radial_Green_matrix}) can be utilized directly to evaluate,
for instance, the two--photon cross sections  $\sigma_2$ for a
\textit{non--resonant} excitation, ionization, or decay process. They also 
occur rather naturally in the theory of the photon scattering on hydrogen--like
ions. In the following, we briefly outline the perturbative calculation of the 
two--photon ionization cross section for hydrogen--like ions which, 
for an unpolarized target and in atomic units%
\footnote{The cross section $\sigma_2$ has the dimension 
$\text{length}^4 \times  \text{time}$ and, thus, can be converted into 
cgs--units cm$^4 \cdot$s by using the multiplication 
factor $1.896792\cdot 10^{-50}$. 
\label{r:units}},
is given by
\begin{eqnarray}
\label{two-photon-sigma}
   \sigma_2 & = & \frac{8\,\pi^3 \,\alpha^2}{E_{\gamma}^{\,2}} \:
   \sum_{\kappa_f m_f} \, \frac{1}{2j_i+1} \: \sum_{m_i}|M_{fi}|^2 \, , 
\end{eqnarray}
where $E_{\gamma}$ is the photon energy and $M_{fi}$ the two--photon 
transition amplitude
\begin{eqnarray} 
\label{tpme} 
   M_{fi} & = & \sum\mkern-26mu\int_{\nu} \; 
   \frac{\langle\psi_f \,|\, \bu_{\lambda_2}\,\expp^{i\bk_2\br}\cdot \bp
                       \,|\, \psi_{\nu}\rangle \:
         \langle\psi_{\nu} \,|\, \bu_{\lambda_1}\, \expp^{i\bk_1\br}\cdot \bp
                           \,|\, \psi_i\rangle
       }{E_{\nu} - E_{\gamma} - E_i} \: . 
\end{eqnarray}
In this amplitude, moreover, $(\psi_i,\, E_i), \: (\psi_{\nu},\, E_{\nu})$,
and $(\psi_{f},\, E_f)$ denote the wave functions and energies of the initial, 
intermediate and final atomic states, respectively. Here, the energy of the 
final state, $E_f$, does not appear explicitly in (\ref{tpme}) but 
follows from 
$$E_f \,=\, E_i \,+\, 2E_{\gamma} \, $$
due to the conservation of energy. Furthermore, the two vector quantities 
$\bu_{\lambda}$ and $\bp$ in the transition amplitude (\ref{tpme}) refer to 
the polarization of the two photons as well as to the electron momentum 
operator.

\medskip

As mentioned before, the summation over $\nu$ in (\ref{tpme}) runs over the
\textit{complete spectrum} of the atom including the continuum. This summation
can be replaced, therefore, by a single Green's function 
(\ref{greens-expansion}), so that the transition amplitude (\ref{tpme}) 
finally takes the form
\begin{eqnarray}
\label{Mp-with-Greens}
   M_{fi} & = &
   \int \psi_f^{\dagger}(\br)\, \bu_{\lambda_2} \,\expp^{i\bk_2\br} \cdot \bp\:
        \rG_{E_i+E_{\gamma}}\,(\br,\br')\: 
        \bu_{\lambda_1}\,\expp^{i\bk_1\br'} \cdot \bp'\, \psi_i(\br') \, 
        d\br\, d\br'\; .
\end{eqnarray}
It is this form of the transition amplitude which has often been used in the
literature to study non--resonant, two--photon processes 
\cite{Maquet:98,Series}.

\subsubsection{Nonrelativistic ionization cross sections} 
 
For the sake of brevity, let us restrict ourselves to the two--photon 
ionization cross sections within the \textit{long--wavelength} approximation, 
i.e.\ we assume $\expp^{i\bk\cdot\br}\,\equiv \,1$ for the coupling of the 
radiation field in (\ref{Mp-with-Greens}). 
Apart from the electric--dipole field, of course, this 
approximation neglects the contribution from all higher multipoles, but is 
known to describe well the ionization of light atoms with a nuclear charge 
of, say, $Z\,\lesssim\,30$ and for photon energies below the ionization 
threshold $E_{\gamma}<E_{\,T}$. By substituting $\bp\rightarrow \br$
and $E_{\gamma} \rightarrow 1/E_{\gamma}$ into Eqs.\  (\ref{two-photon-sigma})
and (\ref{tpme}), moreover, we may obtain the ionization cross section in 
\textit{length gauge} 
\begin{eqnarray}
\label{two-photon-sigma-nr}
   \sigma_2^{\rm (length)} & = & 8 \,\pi^3\, \alpha^2 \, E_{\gamma}^{\,2} \; 
   \sum_{l_f m_f} \,\frac{1}{2l_i+1}\, \sum_{m_i} 
   |M_{fi}^{\rm (length)}|^{\,2},
\end{eqnarray}
with 
\begin{eqnarray}
\label{Mr-with-Greens-nr}
   M_{fi}^{\rm (length)} & = &
   \int \psi_f^{\dagger}\,(\br)\, \bu_{\lambda_2}  \cdot \br \:
   \rG_{E_i+E_{\gamma}}\,(\br,\br')\: \bu_{\lambda_1} \cdot \br' \,
   \psi_i\,(\br') \, d\br\, d\br' \, .
\end{eqnarray}
Using the \textit{radial--angular} representations (\ref{psinlm})
and (\ref{CoulombGreensNonSplit}) of the wave and Green's functions,
respectively, and by making use of some angular momentum algebra, the
6--dimensional integral in the transition amplitude (\ref{Mr-with-Greens-nr}) 
can be reduced further to just a \textit{two--dimensional integration} over the 
radial coordinates $r$ and $r'$. 
In addition, if we assume the ion initially in its $1s$ ground--state
and circular polarized light, i.e.\ two photons with the same
helicity $\lambda_1 \,=\, \lambda_2 \,=\, \pm\,1$, the two--photon ionization
cross section (in length gauge) simply takes the form
\begin{eqnarray}
\label{two-photon-sigma_nr_circ}
   \sigma_2^{\rm (length,\,circular)} & = & 8\pi^3 \alpha^2 \, E_{\gamma}^2 \,
   \left| \int P_{E_f2}\,(r) \: r \: \rg_{E_i+E_{\gamma}, 1}\,(r,r') 
                \:r'\: P_{10}\,(r') \, dr\, dr' \right|^2 \, .
\end{eqnarray}

\subsubsection{Relativistic two--photon ionization cross sections} 

The \textit{long--wavelength} approximation for the coupling of the
radiation field can be considered also within the framework of the 
relativistic theory. In this framework, however, an useful estimate of the 
total cross section $\sigma_2$ are obtained only if the  photon energy is 
well below the threshold energy $E_{\gamma} \,<\, E_{\,T}$ of the two--photon 
ionization. In the relativistic theory, the (long--wavelength) transition 
amplitude (\ref{Mp-with-Greens}) takes the form
\begin{eqnarray}
\label{tpme_lw} 
   M_{fi} & = & 
   c^{\,2} \, 
   \int \, \psi_f^{\dagger}(\br) \,\bu_{\lambda_2} \cdot \balpha \,  
           \rG_{E_i+E_{\gamma}}\,(\br, \br')\,\bu_{\lambda_1}\cdot\balpha'\,
           \psi_i(\br') \, d\br \, d\br', 
\end{eqnarray}
where $\balpha$ denotes Dirac's velocity operator. Using the
radial--angular representation (\ref{greens_radial_Green_matrix}) of 
the Green's functions, then the total two--photon ionization cross section 
$\sigma_2$ for circular--polarized light can be written as
\begin{small}
\begin{eqnarray} 
\label{long_wave_relativistic_circ} 
   \sigma_2^{\rm (velocity,\,circular)} 
   & = & \frac{8\pi^3 }{\alpha^2 \, E_{\gamma}^2} \,
   \left\{ \frac{32}{25} \, U^{\, SL} 
           \left( d_{\frac{5}{2}}, p_{\frac{3}{2}}, s_{\frac{1}{2}} 
           \right)^{\,2}
   \: + \: \frac{12}{2025} \, 
           \left[ 5 \, U^{\,LL}
           \left( d_{\frac{3}{2}}, p_{\frac{3}{2}}, s_{\frac{1}{2}}
           \right) \:+\: \right.\right.
   \nonumber \\[0.4cm]
   &   & \hspace*{-0.9cm} \left. \left. 
           \: + \: 3 \, U^{\,SL}
           \left( d_{\frac{3}{2}}, p_{\frac{3}{2}}, s_{\frac{1}{2}}
           \right) 
           \: - \: 5 \, U^{\,LL} 
           \left( d_{\frac{3}{2}}, p_{\frac{1}{2}}, s_{\frac{1}{2}}
           \right) 
           \: - \: 15 \, U^{\,LS}
           \left( d_{\frac{3}{2}}, p_{\frac{1}{2}}, s_{\frac{1}{2}}
           \right) \right]^{\,2} \right\} ,
\end{eqnarray}  
\end{small}

where we introduced the radial integral
\begin{eqnarray}
\label{radial_lw_me} 
   U^{\,TT'}(\kappa_f, \kappa_{\nu}, \kappa_i) & = & 
   \int \, \rg_{\, E_f\kappa_f}^{\,\overline T}\,(r) \,
           \rg^{\,TT'}_{\,E_i+E_{\gamma},\kappa_{\nu}}\,(r,\,r') \,
           \rg_{\,n_i \kappa_i}^{\,\overline{T}'}\,(r') \, dr \,  dr' \; .
\end{eqnarray}
In this integral, a superscript $\overline{T}$ refers to the conjugate of
$T$, i.e.\ $\overline{T} \,=\, S$ for $T \,=\, L$ and \textit{vice versa},
and $\rg_{\,n\kappa}^{\,L} (r)$ and $\rg_{\,n\kappa}^{\,S} (r)$ are used
to denote the large and small components of the radial spinor 
(\ref{greens_radial_spinor_n}). This notation allows for a very compact 
representation of the multi--photon transition amplitudes which can be applied
also well beyond the long--wavelength approximation.

\medskip

In the \textsc{Greens} library, the procedure 
\procref{greens\_{}two\_{}photon\_{}cs()}
is presented to compute two--photon ionization cross sections in various
approximations.

\subsection{Special functions} 
\label{s:spec-funct}

Of course, the main emphasize in developing the \textsc{Greens} library 
has been paid to the computation of the Coulomb wave and Green's functions 
as appropriate for a theoretical description of hydrogen--like ions.
As seen from sections 2.1 and 2.2, however, for an explicit 
representation of these functions we usually need to refer to a few 
\textit{special functions} such as the $\Gamma (z)$ and $\Psi (z)$ functions, 
or the Kummer and Whittaker functions of the first and second kind which are 
known from the mathematical literature \cite{Abramowitz}. Therefore, in order
to facilitate the implementation of the Coulomb functions, we have to provide 
also a simple interface to these special functions; in the following,
we briefly summarize the definition of these functions and for which type 
of arguments they are needed for the \textsc{Greens} library.

\medskip

Euler's Gamma function $\Gamma (z)$ and the Psi--function $\Psi (z)$
occur very frequently and in quite different fields of physics. While the
$\Gamma -$function is defined by the integral
\begin{eqnarray}
\label{Gamma}
   \Gamma (z) & = & \int_0^{\infty} \, t^{z-1}\,\expp^{-t}\,dt
\end{eqnarray}
the $\Psi -$function refers to the derivative 
\begin{eqnarray}
\label{Psi}
   \Psi(z) & = & \frac{d\,[ln \Gamma(z)]}{dz} \, . 
\end{eqnarray} 
These functions are defined for all \textit{complex arguments} $z$ except of
the real negative integers $z \,\ne\, -1,\,-2,\, ... \,$ where they 
have their poles. In \textsc{Greens}, the $\Gamma (z)$ 
function with real arguments $z$ is needed for the computation of the 
bound--state wave and Green's functions, respectively, while complex arguments 
arise in the representation of the free--electron waves (\ref{radial-PEl}).
The $\Psi-$function, in addition, arises in the calculation of the Kummer
function $U(a,b;z)$ of the second type if the argument $b$ refers to an
integer in the computation of nonrelativistic Green's functions.

\medskip

Although the Coulomb wave and Green's functions are often expressed in terms 
of the Whittaker functions $\rM_{\,a,b}\,(z)$ and $\rW_{\,a,b}\,(z)$ of the 
first and second kind, in practical computations one makes better use
of the Kummer functions of the corresponding kind, as discussed in subsection
2.2.1 above. The Kummer functions $M(a,b;z)$ and $U(a,b;z)$ of the first and
second kind refer to the \textit{regular} and  \textit{irregular}
solutions of Kummer's equation
\begin{eqnarray} 
\label{KummersEquation} 
   z\,\frac{d^{\,2}\,\rM}{dz^{\,2}} \:+\: 
   (b-z)\,\frac{d\rM}{dz} \:-\: a\,\rM & = & 0 \, ;
\end{eqnarray} 
in the literature, however, also several other notations are used for these 
functions such as $\rM(a,b;z) \:=\: _1F_1(a;b;z)$ or 
$\rU(a,b,z) \:=\: \Psi(a,b,z)$, respectively. Usually, the function 
$M(a,b;z)$ of the first kind is solved for the initial value 
$\rM(a,b;0)\:=\:1$ and, hence, is given by the confluent hypergeometric 
series
\begin{eqnarray} 
\label{KummerM} 
   \rM(a,b;z) \:=\: 1 \:+\:
   \frac{a}{b}\,z     \:+\:
   \frac{1}{2}\,\frac{a(a+1)}{b(b+1)}\,z^2 \:+\: \cdots \, .
\end{eqnarray} 
The Kummer function of the first kind $\rM(a,b;z)$ is needed for both, real 
$a,\,b,\, z$ and complex arguments $a,\,z$ to represent the radial wave and 
Green's function components. In contrast, the Kummer function $\rU(a,b;z)$ 
of the second kind is required only for real argument $b$, for which it can 
be expressed as a linear combination
\begin{eqnarray} 
\label{KummerU} 
   \rU(a,b;z) & = & \frac{\pi}{\sin{\pi b}} \,
   \left[\frac{\rM(a,b;z)}{\Gamma(1+a-b)\,\Gamma(b)} 
   \;-\; z^{1-b}\:\frac{\rM(a+1-b, 2-b; z)}{\Gamma(a)\,\Gamma(2-b)}\right]
\end{eqnarray} 
of two Kummer functions of the first kind; the function $\rU(a,b;z)$ 
arises in the computation of the radial Green's function. 

\medskip

The following section explains how these \textit{special functions} from the 
\textsc{Greens} library can be used also in applications other than the 
computation of Coulomb wave and Green's functions.

\section{Program organisation} 
\label{s:lib-descr}

\subsection{Overview about the {\sc Greens} library} 
\label{overview-of-functions}

The \textsc{Greens} library has been designed mainly in order to facilitate 
numerical applications of the Coulomb wave and Green's functions from section 2. 
It provides the user with a  set of C++ procedures to compute the 
\textit{radial} components of these functions within both, a 
\textit{nonrelativistic} as well as \textit{relativistic} framework. 
Apart from the radial components, however, we also support the numerical 
integration of the Coulomb functions as well as the computation of a few 
selected matrix elements which, below, will help us demonstrate the application
of the \textsc{Greens} library. To provide the user with a simple access to 
the various functions, the concepts of object--oriented  programming 
such as \textit{structures}, \textit{classes} and \textit{members} as well as
the \textit{overloading} of procedures and operators have been utilized 
carefully.

\medskip

Table 1 lists the \textit{main} procedures of the \textsc{Greens} library
for calculating the energies and radial components of the Coulomb functions.
To simplify the use of the library, the classes 
\texttt{spinor2\_col}, \texttt{spinor2\_raw}, and \texttt{matrix\_2x2} have 
been implemented to describe the radial spinor (\ref{greens_radial_spinor_n}),
its adjunct raw spinor, and the radial Green's matrix 
(\ref{radial-green-matrix}), respectively. The classes \texttt{spinor2\_col}
and \texttt{spinor2\_raw}, for instance, contain each the two members 
\texttt{.L} and 
\texttt{.S} to represent the large and small components of a relativistic
wave function, while the class \texttt{matrix\_2x2} has the four members
\texttt{.LL, .LS, .SL,} and \texttt{.SS} with an obvious meaning. 
The class \texttt{matrix\_2x2}, moreover, also contains the member 
\texttt{.e} which just returns all the four matrix elements together 
within a $2 \times 2$ array.

\medskip

In order to treat the bound-- and free--electron states in a similar way,
the two wave function procedures \procref{green\_{}radial\_{}orbital()} and
\procref{green\_{}radial\_{}spinor()} have been \textit{overloaded}.
For these two procedures, a first \textit{integer} argument $n \,\ge\,1$ 
is used to represent the principal quantum number and to return the 
corresponding bound--state solution, while a (first) argument $E\,>\,0$ of 
type \texttt{double} refers to the kinetic energy of a free--electron state
(in Hartree units). 
As mentioned above, however, this energy $E$ does not include the electron 
rest energy, neither in the nonrelativistic nor relativistic framework.
The two additional procedures \procref{greens\_{}set\_{}nuclear\_{}charge()} 
and \procref{greens\_{}get\_{}nuclear\_{}charge()} from table 1 can be called 
to re--define or to return the current value of the nuclear charge which 
is utilized for the computation of all radial functions. The default 
value of the nuclear charge is $ Z \,=\, 1$.

\begin{sidewaystable}
\begin{small}
{\bf Table 1}
\hspace{0.2cm}
{\rm Main procedures of the \textsc{Greens} library to calculate the
energies and radial wave and Green's functions for hydrogen--like ions.
The (expected) type of parameters is shown by using the syntax of C++;
all quantities below must be given in atomic units.}
\begin{center}
\begin{tabular}{p{4.2cm} p{7.0cm} p{12.3cm}} \\[-0.6cm]    
\hline \hline  \\[-0.4cm]
Procedure                           & Arguments & Description and comments 
\\[0.1cm]  \hline  \\[-0.2cm]
\texttt{double greens\_energy}      & \texttt{(int n)}   & 
Returns the nonrelativistic energy $E_n$ (in a.~u.) of a bound--state
solution with principal quantum number $n$; see Eq.\ (\ref{energy_non}).      
                                    \\
                                    & \texttt{(int n, int kappa)}   &  
Returns the relativistic energy $E_{n\kappa}$ (in a.~u.) of a 
bound--state solution quantum numbers $n$ and $\kappa$; 
see Eq.\ (\ref{energy_rel}).        \\
\texttt{double \newline greens\_radial\_orbital} 
                                    & \texttt{(int n, int l, double r)}   &
Computes the value of the radial function $P_{nl}(r)$ at $r$ of a bound 
state (\ref{radial-Pnl}) with principal quantum number $n$ and orbital 
angular momentum $l$. 
                                    \\
                                    & \texttt{(double E, int l, double r)} &
Computes the value of the radial function $P_{El}(r)$ at $r$ of a 
free--electron state (\ref{radial-PEl}) with energy $E \,>\,0$ and orbital 
angular momentum $l$. 
                                    \\
\texttt{spinor2\_{}col \newline greens\_radial\_spinor}     
                                    & \texttt{(int n, int kappa, double r)} &
Computes the value of the radial spinor function 
$\begin{pmatrix} P_{n\kappa}\,(r) \\[0.1cm] Q_{n\kappa}\,(r)\end{pmatrix}$
at $r$ of a bound state (\ref{P-component}--\ref{Q-component}) with principal 
quantum number $n$ and relativistic angular momentum  quantum number $\kappa$. 
                                    \\
                                    & \texttt{(double E, int kappa, double r)} &
Computes the value of the radial spinor function 
$\begin{pmatrix} P_{E\kappa}\,(r) \\[0.1cm] Q_{E\kappa}\,(r)\end{pmatrix}$
at $r$ of a free--electron state with energy $E$  
and relativistic angular momentum  quantum number $\kappa$. 
                                    \\
\texttt{double \newline greens\_radial\_function} 
   & \texttt{(double E, int l, 
                                             double r, \newline double r')} &
Computes the radial Coulomb Green's function $\rg_{El}\,(r,\,r')$ at $r$ and
$r'$ from (\ref{radial_Green_function_algebraic}) for the energy $E \,<\,0$ 
and orbital angular momentum $l$.   \\ 
\texttt{matrix\_{}2x2 \newline greens\_radial\_matrix}   
                                    & \texttt{(double E, int kappa, 
                                               double r, \newline double r')} &
Computes the radial Coulomb Green's matrix  
$ \displaystyle\begin{pmatrix} 
                  \rg_{E\kappa}^{\,LL} &  \rg_{E\kappa}^{\,LS} \\[0.1cm] 
                  \rg_{E\kappa}^{\,SL} &  \rg_{E\kappa}^{\,SS}  
                  \end{pmatrix} $  
from (\ref{radial-green-matrix}) at $r$ and $r'$ for the energy 
$E \,<\,0$ and the relativistic angular momentum  quantum number $\kappa$. 
\\[0.1cm]
\hline \hline 
\end{tabular}
\end{center}
\end{small}
\end{sidewaystable}

\begin{sidewaystable}
\begin{small}
{\bf Table 2}
\hspace{0.2cm}
{\rm Utility procedures of the \textsc{Greens} library for the numerical 
integration of 1-- and 2--dimensional functions and the computation
of two--photon ionization cross sections $\sigma_{\,2}$ in various
approximations. The same notation as in table 1 is used.}
\begin{center}
\begin{tabular}{p{4.2cm} p{7.0cm} p{12.3cm}} \\[-0.6cm]    
\hline \hline  \\[-0.4cm]
Procedure                         & Arguments & Description and comments 
\\[0.1cm]  \hline  \\[-0.2cm]
\texttt{double \newline greens\_{}integral\_{}GL}      
   & \texttt{(double(*funct)(double x), double a, \newline double b, int d)}   
   & Calculates the definite (1--dimensional) integral 
     $\int\limits_a^b \, f(x) \, dx$ with an accuracy of (at least) \texttt{d}
     valid digits. This procedure applies an adaptive Gauss--Legendre 
     integration formula, independently in each dimension.          \\
   & \texttt{(double(*funct)(double x), int d)}   
   & Calculates the definite (1--dimensional) integral 
     $\int\limits_0^\infty \, f(x) \, dx$ with an accuracy of (at least) 
     \texttt{d} valid digits if $f(x)$ does not oscillate rapidly and 
     vanishes sufficiently fast for large values of $x$.    \\
   & \texttt{(double(*funct)(double x, double y), \newline 
             double ax, double bx, double ay, \newline double by, int d)}   
   & Calculates the definite (2--dimensional) integral 
     $\int\limits_{a_x}^{b_x} \, \int\limits_{a_y}^{b_y} \, f(x,y) \, dx dy$ 
     with an accuracy of (at least) \texttt{d} valid digits.  \\
   & \texttt{(double(*funct)(double x, double y), \newline int d)}   
   & Calculates the definite (2--dimensional) integral 
     $\int\limits_{0}^{\infty} \, \int\limits_{0}^{\infty} \, f(x,y) \, dx dy$ 
     with an accuracy of (at least) \texttt{d} valid digits if $f(x,y)$ does 
     not oscillate rapidly and vanishes sufficiently fast for large 
     values of $x$ and $y$. 
\\[0.2cm]  
\texttt{double \newline greens\_two\_photon\_cs} 
   & \texttt{("nonrelativistic", "circular", \newline double E\_ph, int d)}   
   & Computes the nonrelativistic two--photon ionization cross section 
     (\ref{two-photon-sigma_nr_circ}) for circular polarized light, in
     long--wavelength approximation, and for a photon energy 
     $E_{\rm ph} \,>\, E_{\rm 1s} / 2$. A cross section value in atomic units 
     and with an accuracy of (at least) \texttt{d} valid digits is returned. \\
   & \texttt{("nonrelativstic", "linear", \newline double E\_ph, int d)}   
   & Computes the nonrelativistic two--photon ionization cross section 
     for linear polarized light and in
     long--wavelength approximation.                  \\   
   & \texttt{("relativistic", "circular", \newline double E\_ph, int d)}   
   & Computes the relativistic two--photon ionization cross section 
     (\ref{long_wave_relativistic_circ}) for circular polarized light, in
     long--wavelength approximation, and for a photon energy 
     $E_{\rm ph} \,>\, E_{\rm 1s} / 2$.                \\
   & \texttt{("relativistic", "linear", \newline double E\_ph, int d)}   &
     Computes the relativistic two--photon ionization cross section 
     for linear polarized light and in
     long--wavelength approximation.  \\[0.1cm]                                   
\hline \hline 
\end{tabular}
\end{center}
\end{small}
\end{sidewaystable}

\begin{table}
\begin{small}
{\bf Table 3}
\hspace{0.2cm}
{\rm Special function procedures of the \textsc{Greens} library. 
The same notation as in table 1 is used. The type of all procedures is
\texttt{double} if all arguments are \texttt{double}, and is of type 
\texttt{complex} otherwise.}
\begin{center}
\begin{tabular}{p{2.2cm} p{6.2cm} p{6.8cm}} \\[-0.4cm]    
\hline \hline  \\[-0.4cm]
Procedure                 & Arguments & Description and comments 
\\[0.1cm]  \hline  \\[-0.2cm]
\texttt{GAMMA}            &  \texttt{(double z)} or
                             \texttt{(complex z)} &
Returns the $\Gamma(z)$ function (\ref{Gamma}) for either a real or complex 
argument $z$.                      \\
 \texttt{Psi}             &  \texttt{(double z)} or 
                             \texttt{(complex z)} &
Returns the $\Psi(z)$ function (\ref{Psi}) for either a real or complex 
argument $z$.                      \\
 \texttt{KummerM}         & \texttt{(double a, double b, double z)} or \newline  
                            \texttt{(complex a, double b, complex z)}    &
Calculates the Kummer function $\rM(a,b;z)$ of the first kind (\ref{KummerM})
for real and/or complex arguments $a,\,b,$ and $z$.  \\
\texttt{KummerU}          & \texttt{(double a, double b, double z)}      & 
Calculates the Kummer function $\rU(a,b;z)$ of the second kind (\ref{KummerU})
for real arguments $a,\,b,$ and $z$.  \\
\texttt{WhittakerM}       & \texttt{(double a, double b, double z)} or \newline
                            \texttt{(complex a, double b, complex z)}  &
Calculates the Whittaker function $\rM_{a,b}(z)$ of the first kind 
(\ref{WhittakerM}) for either real or complex arguments $a,\,b,$ and $z$;
$b$ must be real.  \\
 \texttt{WhittakerW}      & \texttt{(double a, double b, double z)}      & 
Calculates the Whittaker function $\rW_{a,b}(z)$ of the second kind 
(\ref{WhittakerW}) for real arguments $a,\,b,$ and $z$.  \\[0.1cm]
\hline \hline 
\end{tabular}
\end{center}
\end{small}
\end{table}

\medskip

In most applications, the (radial) Coulomb wave and Green's function components
usually occur as part of some matrix element and, hence, first require an 
additional integration (over $r$ and/or $r'$) before any \textit{observable}
quantity is obtained. Therefore, to facilitate such applications, we also 
provide
the utility procedure \procref{greens\_{}integral\_{}GL()} which evaluates a 
1-- or 2--dimensional integral over a finite or infinite area
with a \textit{user--defined} accuracy, see table 2. In this procedure, 
a Gauss--Legendre quadrature \cite{Abramowitz} of appropriate order  is applied,
independently for each dimension of the integrand. Moreover, to ensure
a result which is accurate up to a given number of \texttt{d} valid digits, 
the domain of integration is divided by steps into subdomains until the 
required accuracy is obtained. A \textsc{Warning} arises during the execution, 
if the requested precision cannot be guarranteed by the procedure. As seen from
table 2, the procedure name \procref{greens\_{}integral\_{}GL()} is 
\textit{overloaded} and, thus, can be invoked with 
rather different lists of parameters, from which the dimension of the 
integral, the integration domain as well as the type of the function is deduced. 
Apart from a real--valued integrand
$f(x)$ or $f(x,y)$, respectively, \procref{greens\_{}integral\_{}GL()}
also supports a \texttt{matrix\_{}2x2}--valued integrand as appropriate for the
computation of matrix elements such as (\ref{radial_lw_me}) from the
relativistic theory. In the latter case, for instance, all the four integrals 
$U^{LL}$, $U^{LS}$, $U^{SL}$ and $U^{SS}$ in (\ref{long_wave_relativistic_circ})
could be treated simultaneously.

\medskip

A second utility procedure \procref{greens\_two\_photon\_cs()} from table 2 
enables the user to calculate two--photon ionization cross sections in various
approximations. Obviously, this procedure makes use of 
\procref{greens\_{}integral\_{}GL()} and is mainly provided for test purposes 
below.
It helps compute the total two--photon ionization cross sections $\sigma_2$
for linear or circular polarized light and within either the nonrelativistic
or relativistic framework, respectively. In all of these cases, however, 
the computation of the cross sections is restricted to the long--wavelength 
approximation $\expp^{i\bk\br}\,=\,1$ for the coupling of the radiation field
and to the ionization of an electron from the unpolarized $1s$ ground state.
In addition, the photon energy $E_{\,\gamma}$, i.e.\ the third argument of 
the procedure \procref{greens\_two\_photon\_cs()} must be in the range
$ -E_{\rm 1s}/2 \,<\, E_{\,\gamma} \,<\, -E_{\rm 1s}$ where $E_{\rm 1s}$ 
is the (negative) $1s-$binding energy from Eqs.\  (\ref{energy_non}) or 
(\ref{energy_rel}). Again, the last argument \texttt{d} refers to the 
requested accuracy of the cross section of (at least) \texttt{d} valid digits 
and is transfered directly to the underlying integration procedure 
\procref{greens\_{}integral\_{}GL()}.

\medskip

Of course, the wave and Green's functions from section 2 can hardly be 
implemented without a proper set of \textit{special function} procedures.
Therefore, table 3 displays those procedures which are provided by the 
\textsc{Greens} library and which we briefly discussed in section 2.4. 
The allowed types of the parameters are also displayed in this table.

\subsection{Distribution and compilation of the {\sc Greens} library} 
 
The \textsc{Greens} library will be distributed as the gzipped tar--file
\texttt{greens.tar.gz} from which the \texttt{greens} root directory  is 
obtained by \texttt{gunzip greens.tar.gz} and \texttt{tar -xvf greens.tar}. 
This root contains a 
\texttt{Read.me} file, the \texttt{src} subdirectory for the source code 
as well as six subdirectories for various examples. In \texttt{src}, 
we provide the header file \texttt{greens.h} and a makefile to facilitate 
the compilation of the (static) library \texttt{libgreens.a} in the 
\texttt{greens} root directory. It also incorporates about 50 source files
for all of the individual procedures.

\medskip

In the following section, two examples from the subdirectories 
\texttt{example-coulomb-funct} and \texttt{example-twophoton-cs} are 
discussed in more detail and are taken as the test for the installation 
of the library. Each of these example subdirectories, again, contain a 
makefile from which an executable (\texttt{a.out}) is generated simply by 
typing \texttt{make} within the corresponding subdirectory. Since these 
makefiles also compile and link the library \textbf{libgreens.a}, the 
user may start directly from a copy of one of these subdirectories for his 
own \textit{application} of the \textsc{Greens} library.

\section{Examples} 
\label{examples} 

To illustrate the use of the \textsc{Greens} library, we first show how 
the (radial) Coulomb wave and Green's functions can be calculated for any 
point $r$ or $(r,r')$, respectively. Hereby, a simple comparison between the
nonrelativistic and relativistic theory --- in the limits of a low and high
nuclear charge $Z$ --- is achieved by setting $Z\,=\,1$ (hydrogen) and 
$Z\,=\,92$ (hydrogen--like uranium), respectively. Figure 1 displays
the source code which evaluates the two radial
functions $P_{4d} (r)$ and $P_{4d_{5/2}} (r)$, respectively, for $r-$values in
the range $ r\,=\, 0.,\, \ldots,\, 25./Z$ with a stepsize of 
$\Delta r\,=\, 0.1/Z$. Beside of these wave function components, this code also 
calculates the Coulomb Green's functions
at the same values of $r$ and for a fixed $ r' \,=\, 2.5/Z$. For a call of this
procedure, the printout is (partially) shown in the \textsc{Test Run Output}
below. The source of this example and the complete printout can be found in
the subdirectory \texttt{example-coulomb-funct}. In order to obtain the
---full--- radial part of the Coulomb wave and Green's functions, of course,
the results of \procref{greens\_{}radial\_{}orbital()} and 
\procref{greens\_{}radial\_{}spinor()} must be multiplied with $1/r$, while
the values from \procref{greens\_{}radial\_{}function()} and 
\procref{greens\_{}radial\_{}matrix()} have to be multiplied with $1/rr'$,
respectively.

\begin{figure}
\begin{small}
{\renewcommand\baselinestretch{0.88}
\begin{verbatim}
#include "greens.h"

int main(void){
int    n, l, kappa;              // quantum numbers
double r, rp, E, wf_nr, gf_nr;   // coordinates, energies, etc.
spinor2_col wf_r;                // relativistic spinor
matrix_2x2  gf_r;                // relativistic Green's matrix

print("#Test of the Coulomb radial functions");

for(double Z=1.0; Z<93.0; Z=Z+91.0) {
  print();
  greens_set_nuclear_charge(Z);        // set nuclear charge 
  E = -greens_energy(1) * 0.8;

  rp = 2.5/Z;  n = 4; l = 2; kappa = -3;
  write("# coord        wf_nr        wf_r.L "); 
  print("     gf_nr       gf_r.e[0][0]  gf_r.LL");
  for (r=0.0; r<25.0/Z; r=r+0.1/Z){
    wf_nr = greens_radial_orbital(n, l, r);
    wf_r  = greens_radial_spinor (n, l, r);
    gf_nr = greens_radial_function(-E, l, r, rp);
    gf_r  = greens_radial_matrix  (-E, l, r, rp);
     
  printf("%E %E %E %E %E %E\n", r, wf_nr, wf_r.L, 
                                gf_nr, gf_r.e[0][0], gf_r.LL); }}
return 0;}
\end{verbatim}}
\end{small} 
\vspace*{-0.5cm}
\caption{Calculation of the Coulomb wave and Green's functions for nuclear 
         charge $Z\,=\,1$ and $Z\,=\,92$. The printout of this procedure
         is shown in the \textsc{Test Run Output} and in the file
         \texttt{printout.txt} in the subdirectory 
         \texttt{example-coulomb-funct}.}
\end{figure}

\bigskip

A second example concerns the computation of the two--photon ionization
cross sections for the two ions from above. For these ions, the $1s$ binding
energies are $ -1/2$ and $-4232$ Hartrees within the nonrelativistic
theory. In the \textsc{Test Run Output} below, the two--photon 
ionization cross sections for circular and linear polarized light and within
both, the nonrelativistic and relativistic approximation. For each of these
ions, the cross sections are calculated with an accuracy of about six digits
for the ten energies $E_o,\, E_o + 0.01 * Z^{\,2},\, \ldots,\,
E_o + 0.09 * Z^{\,2} $ where $E_o \,=\, 0.3 * Z^{\,2}$ corresponds to 60 \%{}
of the nonrelativistic $1s$ binding energy. Again, the full source of this
example is provided with the \textsc{Greens} library in the subdirectory 
\texttt{example-twophoton-cs} and, thus, can easily be modified for any other
photon energy.

\section{Summary and outlook} 

To facilitate applications of the 'hydrogen ion model' in quite different 
fields of physics, the \textsc{Greens} library is presented and provides a
set of C++ procedures for the computation of the Coulomb wave and Green's 
functions within both, a nonrelativistic as well as relativistic framework. 
Since C++ is today freely available for most architectures, an
object--oriented approach to the Coulomb problem could be realized without 
the need for special compilers or other mathematical libraries. Apart from 
the radial Coulomb functions, however, \textsc{Greens} also provides a set 
of special functions as well as a few utility procedures to evaluate, for
instance, the two--photon ionization cross sections in long--wavelength
approximation.

\medskip

In the future, various extensions of the \textsc{Greens} library might be 
of great interest for the physics community. Owing to the current design of
several \textit{free--electron laser} (FEL) facilities worldwide, 
for example, systematic investigations on multiphoton processes become more and
more likely also in the EUV and x--ray region, where the inner--shell electron
get involved. For such investigations, which will consider also many--electron
atoms and ions, the generation of \textit{effective} one--particle Green's
functions are certainly desirable. First steps into this direction, 
including the combination with the well--known \textsc{Ratip} package
\cite{Fritzsche:01}, are currently under work in our group.

%
%
%
%
%
%
%

%
%
%
%
%
%
\newpage
\textsc{\Large Test Run output}

\subsubsection*{A. Computation of the radial Coulomb wave and Green's functions}

\begin{footnotesize}
\begin{verbatim}
#Test of the Coulomb radial functions

#Nuclear charge is changed to 1.000000
# coord        wf_nr        wf_r.L       gf_nr       gf_r.e[0][0]  gf_r.LL
0.000000E+00 0.000000E+00 0.000000E+00 0.000000E+00 0.000000E+00 0.000000E+00
1.000000E-01 6.758392E-06 6.761076E-06 9.861696E-05 9.865164E-05 9.865164E-05
2.000000E-01 5.228909E-05 5.230066E-05 7.642130E-04 7.643481E-04 7.643481E-04
...
2.480000E+01 -2.307241E-01 -2.307229E-01 8.979656E-09 8.980284E-09 8.980284E-09
2.490000E+01 -2.295400E-01 -2.295388E-01 8.243999E-09 8.244579E-09 8.244579E-09

#Nuclear charge is changed to 92.000000
# coord        wf_nr        wf_r.L       gf_nr       gf_r.e[0][0]  gf_r.LL
0.000000E+00 0.000000E+00 0.000000E+00 0.000000E+00 0.000000E+00 0.000000E+00
1.086957E-03 6.482422E-05 4.344695E-04 1.071923E-06 5.170738E-06 5.170738E-06
2.173913E-03 5.015393E-04 1.952599E-03 8.306663E-06 2.311893E-05 2.311893E-05
...
2.706522E-01 -2.201670E+00 -2.088423E+00 8.960868E-11 1.641947E-10 1.641947E-10
2.717391E-01 -2.190133E+00 -2.074839E+00 8.226637E-11 1.512671E-10 1.512671E-10
\end{verbatim}
\end{footnotesize}
\subsection*{B. Computation of two-photon ionization cross sections}

\begin{footnotesize}
\begin{verbatim}
#Test of the two-photon ionisation cross sections
#Digits is changed to 6

#Nuclear charge is changed to 1.000000
# E            cs_nr_c      cs_r_c       cs_nr_l      cs_r_l
3.000000E-01 8.728681E-01 8.727765E-01 5.849625E-01 5.849002E-01
3.100000E-01 8.819793E-01 8.818399E-01 5.889291E-01 5.888364E-01
3.200000E-01 9.143732E-01 9.143980E-01 6.095973E-01 6.096138E-01
...
3.800000E-01 5.778480E+00 5.794808E+00 4.829547E+00 4.843563E+00
3.900000E-01 1.792990E-01 1.796899E-01 2.330477E-01 2.333333E-01

#Nuclear charge is changed to 92.000000
# E            cs_nr_c      cs_r_c       cs_nr_l      cs_r_l
2.539200E+03 1.439533E-12 6.927729E-13 9.647196E-13 4.629667E-13
2.623840E+03 1.454559E-12 6.763950E-13 9.712612E-13 4.510878E-13
2.708480E+03 1.507983E-12 6.588038E-13 1.005347E-12 4.392479E-13
...
3.216320E+03 9.529863E-12 6.744809E-13 7.964884E-12 4.646710E-13
3.300960E+03 2.956996E-13 7.498654E-13 3.843421E-13 5.232579E-13
\end{verbatim}
\end{footnotesize}

\end{document}